\documentclass[]{revtex4}
\usepackage{amsfonts}
\usepackage{amssymb}
\usepackage{amsmath}

\begin{document}
\title{Improved method for partial-wave decomposition of two-pion exchange three-nucleon force}
\author{Rimantas Lazauskas}
\email{rimantas.lazauskas@ires.in2p3.fr}
\affiliation{IPHC, IN2P3-CNRS/Universit\'e Louis Pasteur BP 28, F-67037 Strasbourg Cedex
2, France}

\begin{abstract}
In this article an efficient method to calculate the matrix elements of
three-nucleon force is presented. The new method is improved version of
partial-wave decomposition of ref.~\cite{Huber_3NF}, which simplifies
expression to be evaluated and permits to reduce computational effort as much
as six times. Proposed method naturally applies to Faddeev-type
calculations but also can be used by any method relaying on partial-wave
decomposition.
\end{abstract}
\maketitle

\section{Introduction}

Three nucleon force (3NF) becomes an indispensable ingredient in realistic
atomic nucleus calculations. In fact discussion is still vivid about the
origin and structure of dominant three-nucleon force terms but even largest
sceptics recognize its importance~in nuclear physics~\cite{3NF_gen}.
Complexity of three-nucleon force turns to be a major obstacle towards its
better understanding. Numerical implementation requires enormous analytical
and programming effort. At the same time even the best algorithms evaluating
3NF matrix elements become huge computer resources demanding tasks. In ref.~%
\cite{Huber_3NF} a nice Partial Wave Decomposition (PWD) technique of 3NF
originating from two-pion exchange diagrams has been demonstrated. The
present article improves this PWD by helping to simplify numerical
implementation procedure as well as reducing number of necessary floating
point operations.

\section{General remarks about PWD technique}

Three-nucleon force naturally decompose into three parts, which are
identical up to a cyclic or anticyclic permutation of the three interacting
particles:
\begin{equation}
V_{ijk}=V_{jk}^{(i)}+V_{ki}^{(j)}+V_{ij}^{(k)},  \label{3NF_pot}
\end{equation}%
here in term $V_{ij}^{(k)}$ particle $k$ is considered as a formal
spectator. While three-nucleon force $V_{ijk}$ is symmetric for exchange of
any two nucleons, term $V_{ij}^{(k)}$ is symmetric only for interchange of
particles $i$ and $j$. When working in isospin formalism,- since systems
wave function $\Psi $ is fully antisymmetric,- it is necessary to evaluate
only one of three terms to get expectation value of the 3NF acting between
three particles:
\begin{equation}
\left\langle \Psi \left\vert V_{ijk}\right\vert \Psi \right\rangle
=3\left\langle \Psi \left\vert V_{ij}^{(k)}\right\vert \Psi \right\rangle .
\label{Exp_val_Gl}
\end{equation}%
Faddeev-type equations also require evaluation of single 3NF term, if they
are written as follows:
\begin{equation}
F_{ij}^{(k)}=G_{0}V_{ij}\Psi +G_{0}V_{ij}^{(k)}\Psi ,  \label{Fad_eq_Gl}
\end{equation}%
here $F_{ij}^{(k)}$ is so called Faddeev component, $V_{ij}$ is two-nucleon
interaction potential between nucleons $i$ and $j$. Faddeev components are
not fully antisymmetric, however they are antisymmetric for interchange of
the nucleons in the active pair (i.e. component $F_{ij}^{(k)}$ is
antisymmetric for exchange of nucleons $i$ and $j$). Fully antisymmetric
systems wave function is a sum of three Faddeev components $\Psi
=F_{jk}^{(i)}+F_{ki}^{(j)}+F_{ij}^{(k)}$.

It is useful to work with Jacobi coordinates or momenta. One has three
different Jacobi momenta sets, each selecting one of three particles as a
spectator. I.e. if we select particle $\left( k\right) $ as spectator,
associated Jacobi momenta are defined by $\overrightarrow{p}_{k}=\frac{1}{2}(%
\overrightarrow{k}_{j}-\overrightarrow{k}_{i})\smallskip $ and $%
\overrightarrow{q}_{k}=(\frac{2}{3}\overrightarrow{k}_{k}-\frac{%
\overrightarrow{k}_{i}+\overrightarrow{k}_{j}}{3})$, where $k_{i}$, $k_{j}$
and $k_{k}$ are individual momenta of the particles. This momenta set is
proper to Fadeev component $F_{ij}^{(k)}$. Following the standard procedure
each Faddeev component $F_{ij}^{(k)}$ is expanded in its proper bipolar
harmonic basis $\left\vert pq\alpha \right\rangle _{k}$:
\begin{equation}
\left\vert pq\alpha \right\rangle _{k}\equiv \left\vert
p_{k}q_{k}\right\rangle \left\vert \left( l_{p}\left( s_{i}s_{j}\right) \!\!{%
\atop{}{{}{{s_{p}}}}}\right) \!\!{\atop{}{{}{j_{p}}}}\left(
l_{q}s_{k}\right) \!\!{\atop{}{{}{j_{q}}}}\right\rangle _{JM}\left\vert
\left( t_{i}t_{j}\right) \!\!{\atop{}{{}{t_{p}}}}t_{k}\right\rangle
_{TT_{z}},  \label{EQ_FA_exp}
\end{equation}%
here index $\alpha $ spans all the symmetry allowed combinations of the
quantum numbers presented in the brackets: $l_{p}$ and $l_{q}$ are the
partial angular momenta associated with respective Jacobi momenta; $s_{i}$
and $t_{i}$ are the spins and isospins of the individual nucleons. Three
nucleon system conserve total angular momentum $J$ and its projection $M.$
Individual isospins of the three nucleons couple to the total isospin $T$
and its magnetic quantum number $T_{z}$.

Separation of the fully symmetric three-nucleon force into symmetric binary
terms, as shown in eq.(\ref{3NF_pot}), can be done in infinitely many
different ways. In practice one associates the binary term $V_{ij}^{(k)}$
with an expression obtained from the Feynman diagram where nucleon $k$ is
chosen as a formal spectator. Three-nucleon force related with two-pion
exchange diagrams in its general form is usually written as~\cite{Epel_3NF}:
\begin{eqnarray}
V_{ij}^{(k)} &=&\frac{\left( \overrightarrow{\sigma }_{i}\centerdot%
\overrightarrow{Q}_{i}\right) \left( \overrightarrow{\sigma }_{j}\centerdot%
\overrightarrow{Q}_{j}\right) }{\left( \overrightarrow{Q}_{i}^{2}+M_{\pi
}^{2}\right) \left( \overrightarrow{Q}_{j}^{2}+M_{\pi }^{2}\right) }%
f_{i}f_{j}\times  \label{2pi_force} \\
&&\left[ C_{1}\left( \overrightarrow{\tau }_{i}\centerdot\overrightarrow{%
\tau }_{j}\right) +C_{3}\left( \overrightarrow{\tau }_{i}\centerdot%
\overrightarrow{\tau }_{j}\right) \left( \overrightarrow{q}_{i}\centerdot%
\overrightarrow{q}_{j}\right) +C_{4}\left( \overrightarrow{\sigma }%
_{k}\centerdot\overrightarrow{Q}_{i}\times \overrightarrow{Q}_{j}\right)
\left( \overrightarrow{\tau }_{k}\centerdot\overrightarrow{\tau }_{i}\times
\overrightarrow{\tau }_{j}\right) \right]  \notag
\end{eqnarray}%
where $C_{1},C_{3}$ and $C_{4}$ are simply multiplicative constants and $%
f_{i}$ is scalar regularization function intended to cut out high-momenta
contribution of the nucleon $i$; there is no strict regularization procedure
to define cut-off function $f_{i}$ and several different regularization
procedures have been used~\cite{Epel_3NF,TM_3NF,Navr_3NF}.

The last expression can be written in a compact form as:
\begin{equation}
V_{ij}^{(k)}=C_{1}W_{a}^{(i)}W_{a}^{(j)}I^{(k)}+C_{3}(\overrightarrow{W}%
_{b}^{(i)}\centerdot\overrightarrow{W}_{b}^{(j)})I^{(k)}+C_{4}(^{(k)}%
\overrightarrow{W}_{d}^{(j)}\centerdot\overrightarrow{W}_{b}^{(i)})I_{0}
\label{2pi_smp}
\end{equation}%
with $I^{(k)}=\left( \overrightarrow{\tau }_{i}\centerdot\overrightarrow{%
\tau }_{j}\right) $, $I_{0}=\left( \overrightarrow{\tau }_{k}\centerdot%
\overrightarrow{\tau }_{i}\times \overrightarrow{\tau }_{j}\right) $, $%
W_{a}^{(i)}=\widetilde{f}_{i}\left( \overrightarrow{\sigma }_{i}\centerdot%
\overrightarrow{Q}_{i}\right) $, $\overrightarrow{W}_{b}^{(i)}=\widetilde{f}%
_{i}\left( \overrightarrow{\sigma }_{i}\centerdot\overrightarrow{Q}%
_{i}\right) \overrightarrow{Q}_{i}$, $^{(k)}\overrightarrow{W}_{d}^{(j)} =%
\widetilde{f}_{j}\left( \overrightarrow{\sigma }_{j}\centerdot%
\overrightarrow{Q}_{j}\right) \left[ \overrightarrow{Q}_{j}\times
\overrightarrow{\sigma }_{k}\right] $ and $\widetilde{f}_{i}=f_{i}/\left(
\overrightarrow{Q}_{i}^{2}+M_{\pi }^{2}\right) $. The momentum transfer $%
\overrightarrow{Q}_{i}=\overrightarrow{k}_{i}^{\prime }-\overrightarrow{k}%
_{i}$ is proper to Jacobi momenta set, where nucleon $i$ is considered as
spectator, giving: $\overrightarrow{Q}_{i}=\overrightarrow{q}_{i}^{\prime }-%
\overrightarrow{q}_{i}$.

In the following we will omit multiplicative constants $C$. Then all three
terms of eq.(\ref{2pi_smp}) applied on the vector $\Psi $ regain the
following form:
\begin{equation}
\Psi ^{\prime }\sim W^{(i)}I\,W^{(j)}\Psi .  \label{2pi_smp_vec}
\end{equation}

H\"{u}ber et al.~\cite{Huber_3NF} proposed a nice method to evaluate these
terms using the following expression:
\begin{eqnarray}
\left. _{k}\left\langle p^{\prime \prime \prime \prime }q^{\prime \prime
\prime \prime }\alpha ^{\prime \prime \prime \prime }\right. \left\vert \Psi
^{\prime }\right\rangle \right. &=&\sum \!\!\!\!\!\!\!\!\int^{\prime \prime
\prime }\sum \!\!\!\!\!\!\!\!\int^{\prime \prime }\sum
\!\!\!\!\!\!\!\!\int^{\prime }\frac{(-)^{J-J^{\prime }}}{\widehat{J}}\left.
_{k}\left\langle p^{\prime \prime \prime \prime }q^{\prime \prime \prime
\prime }\alpha ^{\prime \prime \prime \prime }\right. \left\vert p^{\prime
\prime \prime }q^{\prime \prime \prime }\alpha ^{\prime \prime \prime
}\right\rangle _{j}\right.  \notag \\
&&\times \left. _{j}\left\langle p^{\prime \prime \prime }q^{\prime \prime
\prime }\alpha ^{\prime \prime \prime }\right\Vert W^{(i)}\left\Vert
p^{\prime \prime }q^{\prime \prime }\alpha ^{\prime \prime }\right\rangle
_{j}\right. \left. _{j}\left\langle p^{\prime \prime }q^{\prime \prime
}\alpha ^{\prime \prime }\right\vert I\left\vert p^{\prime }q^{\prime
}\alpha ^{\prime }\right\rangle _{i}\right.  \notag \\
&&\times \left. _{i}\left\langle p^{\prime }q^{\prime }\alpha ^{\prime
}\right\Vert W^{(j)}\left\Vert pq\alpha \right\rangle _{i}\right. \left.
_{i}\left\langle pq\alpha \right. \left\vert \Psi \right\rangle \right. ,
\label{2pi_on_vec}
\end{eqnarray}%
here we have introduced convenient abbreviation $\widehat{J}\equiv 2J+1$. In
this expression $J$ is the angular momentum of the three nucleon system and
eventually of the bases $\alpha ^{\prime \prime \prime \prime },\alpha
^{\prime \prime \prime }$ and $\alpha $. If potential term $W$ is defined as
vector ($\overrightarrow{W}_{b}$ and $\overrightarrow{W}_{d}$ cases) the
angular momentum $J^{\prime }$ of the states $\alpha ^{\prime \prime }$ and $%
\alpha ^{\prime }$ can be different from $J$. Last expression permits to
evaluate matrix elements of the potentials $W$ in their natural coordinate
set, where $Q$ is independent of one Jacobi momenta ($p$ or $q$). This makes
evaluation of matrix element $\left. _{i}\left\langle p^{\prime }q^{\prime
}\alpha ^{\prime }\right\Vert W^{(j)}\left\Vert pq\alpha \right\rangle
_{i}\right. $ quite rapid. Nevertheless one requires two Jacobi basis
transformation operations, namely $\left. _{k}\left\langle p^{\prime \prime
\prime \prime }q^{\prime \prime \prime \prime }\alpha ^{\prime \prime \prime
\prime }\right. \left\vert p^{\prime \prime \prime }q^{\prime \prime \prime
}\alpha ^{\prime \prime \prime }\right\rangle _{j}\right. $ and $\left.
_{j}\left\langle p^{\prime \prime }q^{\prime \prime }\alpha ^{\prime \prime
}\right\vert I\left\vert p^{\prime }q^{\prime }\alpha ^{\prime
}\right\rangle _{i}\right. $, which consumes almost all the time required to
estimate 3NF terms. Furthermore all the intermediate states ($\alpha
^{\prime }$,$\alpha ^{\prime \prime }$ or $\alpha ^{\prime \prime \prime }$)
does not retain antisymmetry with respect to interchange of the nucleons in
the active (coupled) pair. Therefore one is obliged to double the size of
intermediate bases by including unphysical symmetric states, which are
comfortably ignored when using Faddeev-type representation of systems
wave-function.

\section{Improved PWD technique}

In this study we propose to redistribute 3NF terms in Faddeev equation, in
the following manner:
\begin{equation}
\widetilde{F}_{ij}^{(k)}=G_{0}V_{ij}\Psi +\frac{1}{2}G_{0}\left[
V_{jk}^{(i)}+V_{ki}^{(j)}\right] \Psi .  \label{Fad_eq_N}
\end{equation}%
Obviously these Faddeev equations satisfy the same three-particle
Hamiltonian as eq.(\ref{Fad_eq_Gl}), thus they provide the same solution for
the systems wave function $\Psi $. Nevertheless individual Faddeev
components $\widetilde{F}_{ij}^{(k)}$ differ from components $F_{ij}^{(k)},$
which satisfy eq.(\ref{Fad_eq_Gl}).

Instead of using eq.(\ref{Exp_val_Gl}), we calculate 3NF expectation value
using identity:
\begin{equation}
\left\langle \Psi \left\vert V_{ijk}\right\vert \Psi \right\rangle =\frac{3}{%
2}\left\langle \Psi \left\vert V_{jk}^{(i)}+V_{ki}^{(j)}\right\vert \Psi
\right\rangle .
\end{equation}
Now eq.(\ref{2pi_smp_vec}) is replaced with:
\begin{equation}
\Psi ^{\prime }\sim \frac{1}{2}\left[ W^{(k)}I^{(j)}\,W^{(i)}+W^{(k)}%
\,I^{(i)}\,W^{(j)}\right] \Psi .
\end{equation}

Sum of first two terms ($C_{1}$ and $C_{3}$ ones) of 2-pion exchange force
in eq.(\ref{2pi_smp}) we propose to evaluate as:
\begin{eqnarray}
\left. _{k}\left\langle p^{\prime \prime \prime }q^{\prime \prime \prime
}\alpha ^{ \prime \prime \prime }\right. \left\vert \Psi ^{\prime
}\right\rangle \right. &{\small =}&\frac{1}{2}\sum
\!\!\!\!\!\!\!\!\int^{\prime \prime }\sum \!\!\!\!\!\!\!\!\int^{\prime }%
\frac{(-)^{J-J^{\prime }}}{\widehat{J}}  \notag \\
&&\times \left( \left. _{k}\left\langle p^{\prime \prime \prime }q^{ \prime
\prime \prime }\alpha ^{ \prime \prime \prime }\right\Vert W^{(k)}\left\Vert
p^{\prime \prime }q^{\prime \prime }\alpha ^{\prime \prime }\right\rangle
_{k}\right. \left. _{k}\left\langle p^{\prime \prime }q^{\prime \prime
}\alpha ^{\prime \prime }\right\vert I^{(j)}\left\vert p^{\prime }q^{\prime
}\alpha ^{\prime }\right\rangle _{i}\right. \right.  \notag \\
&&\times \left. _{i}\left\langle p^{\prime }q^{\prime }\alpha ^{\prime
}\right\Vert W^{(i)}\left\Vert pq\alpha \right\rangle _{i}\right. \left.
_{i}\left\langle pq\alpha \right. \left\vert \Psi \right\rangle \right.
\notag \\
&&{\small +}\left. _{k}\left\langle p^{ \prime \prime \prime }q^{\prime
\prime \prime }\alpha ^{ \prime \prime \prime }\right\Vert W^{(k)}\left\Vert
p^{\prime \prime }q^{\prime \prime }\alpha ^{\prime \prime }\right\rangle
_{k}\right. \left. _{k}\left\langle p^{\prime \prime }q^{\prime \prime
}\alpha ^{\prime \prime }\right\vert I^{(i)}\left\vert p^{\prime }q^{\prime
}\alpha ^{\prime }\right\rangle _{j}\right.  \notag \\
&&\left. \times \left. _{j}\left\langle p^{\prime }q^{\prime }\alpha
^{\prime }\right\Vert W^{(j)}\left\Vert pq\alpha \right\rangle _{j}\right.
\left. _{j}\left\langle pq\alpha \right. \left\vert \Psi \right\rangle
\right. \right) {\small .}  \label{2pi_on_vec1}
\end{eqnarray}%
Since systems wave function is completely antisymmetric, one has $\left.
_{i}\left\langle pq\alpha \right. \left\vert \Psi \right\rangle \right.
\equiv \left. _{j}\left\langle pq\alpha \right. \left\vert \Psi
\right\rangle \right. $. Furthermore $W(\overrightarrow{Q}_{j})$ is
symmetric for exchange of the particles $k$ and $i$, thus product $\left.
_{j}\left\langle p^{\prime }q^{\prime }\alpha ^{\prime }\right\Vert
W^{(j)}\left\Vert pq\alpha \right\rangle _{j}\right. \left. _{j}\left\langle
pq\alpha \right. \left\vert \Psi \right\rangle \right. $ is a vector
antisymmetric for exchange of the particles $k$ and $i$. One can further
demonstrate simple relation between matrix elements of $\left.
_{k}\left\langle p^{\prime \prime }q^{\prime \prime }\alpha ^{\prime \prime
}\right\vert I^{(j)}\left\vert p^{\prime }q^{\prime }\alpha ^{\prime
}\right\rangle _{i}\right. $ and $\left. _{k}\left\langle p^{\prime \prime
}q^{\prime \prime }\alpha ^{\prime \prime }\right\vert I^{(i)}\left\vert
p^{\prime }q^{\prime }\alpha ^{\prime }\right\rangle _{j}\right. $:
\begin{equation}
\left. _{k}\left\langle p^{\prime \prime }q^{\prime \prime }\alpha ^{\prime
\prime }\right\vert I^{(j)}\left\vert p^{\prime }q^{\prime }\alpha ^{\prime
}\right\rangle _{i}\right. =(-)^{t_{p}^{\prime \prime }+s_{p}^{\prime \prime
}+l_{p}^{\prime \prime }}(-)^{t_{p}^{\prime }+s_{p}^{\prime }+l_{p}^{\prime
}}\left. _{k}\left\langle p^{\prime \prime }q^{\prime \prime }\alpha
^{\prime \prime }\right\vert I^{(i)}\left\vert p^{\prime }q^{\prime }\alpha
^{\prime }\right\rangle _{j}\right. ,
\end{equation}%
where two phase factors reflect the symmetry of bra and ket states with
respect to exchange of two active particles. Second phase factor is
negative, since ket state is antisymmetric for $\left( ki\right) $ pair.
Furthermore matrices $\left. _{i}\left\langle p^{\prime }q^{\prime }\alpha
^{\prime }\right\vert W^{(i)}\left\vert pq\alpha \right\rangle _{i}\right. $
and $\left. _{j}\left\langle p^{\prime }q^{\prime }\alpha ^{\prime
}\right\vert W^{(j)}\left\vert pq\alpha \right\rangle _{j}\right. $ are
identical. Then it is easy to see that two terms in parentheses of eq.(\ref%
{2pi_on_vec1}) cancel each other, unless the $\left\vert p^{\prime \prime
}q^{\prime \prime }\alpha ^{\prime \prime }\right\rangle _{k}$ states are
antisymmetric for exchange of the particles $i$ and $j$ (as should be, since
$V_{jk}^{(i)}+V_{ki}^{(j)}$ as well as $W^{(k)}$ are symmetric in $\left(
ij\right) $ pair). When $\left\vert p^{\prime \prime }q^{\prime \prime
}\alpha ^{\prime \prime }\right\rangle _{k}$ are physical, antisymmetric for
exchange of the particles $i$ and $j$, two terms in eq.(\ref{2pi_on_vec1})
give identical contributions. Finally, one can write:
\begin{eqnarray}
\left. _{k}\left\langle p^{\prime\prime \prime }q^{\prime \prime \prime
}\alpha ^{\prime \prime \prime }\right. \left\vert \Psi ^{\prime
}\right\rangle \right. &{\small =}&\sum \!\!\!\!\!\!\!\!\int^{\prime \prime
}\sum \!\!\!\!\!\!\!\!\int^{\prime }\frac{(-)^{J-J^{\prime }}}{\widehat{J}}%
\left. _{k}\left\langle p^{\prime \prime\prime }q^{\prime \prime \prime
}\alpha ^{\prime \prime \prime }\right\Vert W^{(k)}\left\Vert p^{\prime
\prime }q^{\prime \prime }\alpha ^{\prime \prime }\right\rangle _{k}\right.
\label{Mat_form_me1} \\
&&\times \left. _{k}\left\langle p^{\prime \prime }q^{\prime \prime }\alpha
^{\prime \prime }\right\vert I^{(i)}\left\vert p^{\prime }q^{\prime }\alpha
^{\prime }\right\rangle _{j}\right. \left. _{j}\left\langle p^{\prime
}q^{\prime }\alpha ^{\prime }\right\Vert W^{(j)}\left\Vert pq\alpha
\right\rangle _{j}\right. \left. _{j}\left\langle pq\alpha \right.
\left\vert \Psi \right\rangle \right.  \notag
\end{eqnarray}

This expression contains only one particle-base transformation operation,
which acts only between the physical states (ones being antisymmetric with
respect to active particle pair). On the contrary old PWD eq.(\ref%
{2pi_on_vec}) required\ two particle-base transformation operations, which
were coupling both unphysical (symmetric) and physical (antisymmetric) states.

Now we turn our attention to the $C_{4}$ term of eq.(\ref{2pi_smp}), let
evaluate it according to:
\begin{eqnarray}
\left. _{k}\left\langle p^{\prime \prime \prime }q^{\prime \prime \prime
}\alpha ^{\prime \prime \prime }\right. \left\vert \Psi ^{\prime
}\right\rangle \right. &{\small \sim }&\frac{1}{2}\sum
\!\!\!\!\!\!\!\!\int^{\prime \prime }\sum \!\!\!\!\!\!\!\!\int^{\prime }%
\frac{(-)^{J-J^{\prime }}}{\widehat{J}}  \notag \\
&&\left( \left. _{k}\left\langle p^{\prime \prime \prime }q^{ \prime \prime
\prime }\alpha ^{\prime \prime \prime }\right\Vert \overrightarrow{W}%
_{b}^{(k)}\left\Vert p^{\prime \prime }q^{\prime \prime }\alpha ^{\prime
\prime }\right\rangle _{k}\right. \left. _{k}\left\langle p^{\prime \prime
}q^{\prime \prime }\alpha ^{\prime \prime }\right\vert I_{0}\left\vert
p^{\prime }q^{\prime }\alpha ^{\prime }\right\rangle _{i}\right. \right.
\notag \\
&&\times \left. _{i}\left\langle p^{\prime }q^{\prime }\alpha ^{\prime
}\right\Vert ^{(j)}\overrightarrow{W}_{d}^{(i)}\left\Vert pq\alpha
\right\rangle _{i}\right. \left. _{i}\left\langle pq\alpha \right.
\left\vert \Psi \right\rangle \right.  \notag \\
&&{\small +}\left. _{k}\left\langle p^{\prime \prime \prime }q^{\prime
\prime \prime }\alpha ^{\prime \prime \prime }\right\Vert \overrightarrow{W}%
_{b}^{(k)}\left\Vert p^{\prime \prime }q^{\prime \prime }\alpha ^{\prime
\prime }\right\rangle _{k}\right. \left. _{k}\left\langle p^{\prime \prime
}q^{\prime \prime }\alpha ^{\prime \prime }\right\vert -I_{0}\left\vert
p^{\prime }q^{\prime }\alpha ^{\prime }\right\rangle _{j}\right.  \notag \\
&&\left. \times \left. _{j}\left\langle p^{\prime }q^{\prime }\alpha
^{\prime }\right\Vert ^{(i)}\overrightarrow{W}_{d}^{(j)}\left\Vert pq\alpha
\right\rangle _{j}\right. \left. _{j}\left\langle pq\alpha \right.
\left\vert \Psi \right\rangle \right. \right) {\small .}  \label{2pi_on_vec2}
\end{eqnarray}%
One should note the minus sign in front of the isospin term in the second
line, which appears when splitting $V_{jk}^{(i)}$ into the product $I_{0}(%
\overrightarrow{W}_{b}^{(k)}\centerdot^{(i)}\overrightarrow{W}_{d}^{(j)})$.
The matrix elements of the potentials $^{(j)}\overrightarrow{W}_{d}^{(i)} $
and $\left. ^{(i)}\overrightarrow{W}_{d}^{(j)}\right. $ are equal up to
phase factor:
\begin{equation}
\left. _{i}\left\langle p^{\prime }q^{\prime }\alpha ^{\prime }\right\Vert
^{(j)}\overrightarrow{W}_{d}^{(i)}\left\Vert pq\alpha \right\rangle
_{i}\right. =\left( -\right) ^{s_{p}^{\prime }-s_{p}}\left. _{j}\left\langle
p^{\prime }q^{\prime }\alpha ^{\prime }\right\Vert ^{(i)}\overrightarrow{W}%
_{d}^{(j)}\left\Vert pq\alpha \right\rangle _{j}\right.  \label{Wdd_sym}
\end{equation}
Term $\left. ^{(i)}\overrightarrow{W}_{d}^{(j)}\right. $ is not anymore
symmetric with respect to exchange of the particles $k$ and $i,$ therefore
these matrices also couple physical (antisymmetric) states with the unphysical (symmetric) ones.
Basically, phase factor in eq.(\ref{Wdd_sym}) is positive for the transition
between the physical states and is negative when one pass from physical
state to an unphysical one. For the matrix elements of the isospin operators
one has relation:
\begin{equation}
_{k}\left\langle p^{\prime \prime }q^{\prime \prime }\alpha ^{\prime \prime
}\right\vert I_{0}\left\vert p^{\prime }q^{\prime }\alpha ^{\prime
}\right\rangle _{i} =(-)^{t_{p}^{\prime \prime }+s_{p}^{\prime \prime
}+l_{p}^{\prime \prime }}(-)^{t_{p}^{\prime }+s_{p}^{\prime }+l_{p}^{\prime
}} \left._{k}\left\langle p^{\prime \prime }q^{\prime \prime }\alpha ^{\prime
\prime }\right\vert -I_{0}\left\vert p^{\prime }q^{\prime }\alpha ^{\prime
}\right\rangle _{j}  \right.\label{Iso_eq2}
\end{equation}%
Since sum of potential's terms $V_{jk}^{(i)}+V_{ki}^{(j)}$ as well as term $%
\overrightarrow{W}_{b}^{(k)}$ are symmetric with respect to exchange of the
particles $i$ and $j$ the state $\left\vert p^{\prime \prime }q^{\prime
\prime }\alpha ^{\prime \prime }\right\rangle _{k}$ present in the eq.(\ref%
{2pi_on_vec2}) is antisymmetric for $\left( ij\right) $ pair: thus first
phase factor in eq.(\ref{Iso_eq2}) is negative. Overall phase factor is
negative if $\left\vert p^{\prime }q^{\prime }\alpha ^{\prime }\right\rangle
_{j}$ state is symmetric, whereas overall phase factor is positive if this
state is antisymmetric with respect to pair $(ki)$. Summarizing these
remarks one can see that two additive terms in eq.(\ref{2pi_on_vec2}) are
identical, giving the matrix expression:
\begin{eqnarray}
\left. _{k}\left\langle p^{\prime \prime \prime }q^{\prime \prime \prime
}\alpha ^{\prime \prime \prime }\right. \left\vert \Psi ^{\prime
}\right\rangle \right. &{\small =}&{\small -}\sum
\!\!\!\!\!\!\!\!\int^{\prime \prime }\sum \!\!\!\!\!\!\!\!\int^{\prime }%
\frac{(-)^{J-J^{\prime }}}{\widehat{J}}\left. _{k}\left\langle p^{\prime
\prime \prime}q^{\prime \prime \prime }\alpha ^{\prime \prime \prime
}\right\Vert \overrightarrow{W}_{b}^{(k)}\left\Vert p^{\prime \prime
}q^{\prime \prime }\alpha ^{\prime \prime }\right\rangle _{k}\right. \\
&&\times \left. _{k}\left\langle p^{\prime \prime }q^{\prime \prime }\alpha
^{\prime \prime }\right\vert I_{0}\left\vert p^{\prime }q^{\prime }\alpha
^{\prime }\right\rangle _{j}\right. \left. _{j}\left\langle p^{\prime
}q^{\prime }\alpha ^{\prime }\right\Vert ^{(i)}\overrightarrow{W}%
_{d}^{(j)}\left\Vert pq\alpha \right\rangle _{j}\right. \left.
_{j}\left\langle pq\alpha \right. \left\vert \Psi \right\rangle \right.
\notag
\end{eqnarray}%
This time $\left\vert p^{\prime }q^{\prime }\alpha ^{\prime }\right\rangle
_{j}$ basis contain both physical and unphysical states, nevertheless other
states ($\alpha $, $\alpha ^{\prime \prime }$ and $\alpha ^{\prime \prime
\prime }$) are physical.

\bigskip

Nevertheless one can avoid arithmetics with unphysical states even evaluating
term $C_{4}$. In that aim we use identity $\overrightarrow{\sigma }_{k}=2%
\overrightarrow{S}-\overrightarrow{\sigma }_{i}-\overrightarrow{\sigma }_{j}$
with $\overrightarrow{S}$ representing total spin of three nucleon system to
rewrite $C_{4}$ term:
\begin{equation}
C_{4}(^{(k)}\overrightarrow{W}_{d}^{(j)}\centerdot\overrightarrow{W}%
_{b}^{(i)})I_{0}=C_{4}(\overrightarrow{W}_{e}^{(j)}\centerdot\overrightarrow{%
W}_{b}^{(i)})I_{0}+C_{4}(\overrightarrow{W}_{b}^{(j)}\centerdot%
\overrightarrow{W}_{f}^{(i)})I_{0},
\end{equation}%
where ${\small \overrightarrow{W}_{e}^{(j)}=\widetilde{f}_{j}\left(
\overrightarrow{\sigma }_{j}\centerdot\overrightarrow{Q}_{j}\right) \left[
\overrightarrow{Q}_{j}\times \left( 2\overrightarrow{S}-\overrightarrow{%
\sigma }_{j}\right) \right] }$ and $\overrightarrow{W}_{f}^{(i)}=\widetilde{f%
}_{i}\left( \overrightarrow{\sigma }_{i}\centerdot\overrightarrow{Q}%
_{i}\right) \left[ \overrightarrow{Q}_{i}\times \overrightarrow{\sigma }_{i}%
\right]$.  The particular form of the last operator, as it is easily
demonstrated in the Appendix, results in $\overrightarrow{W}_{f}^{(i)}\equiv0
$. One can see that the term $\overrightarrow{W}_{e}^{(j)}$ is symmetric for
nucleon pair $(ki)$. Therefore  arguments used to derive eq.(%
\ref{Mat_form_me1})  can be followed to demonstrate:
\begin{eqnarray}
\left. _{k}\left\langle p^{\prime \prime \prime }q^{\prime \prime \prime
}\alpha ^{\prime \prime \prime }\right. \left\vert \Psi ^{\prime
}\right\rangle \right. &{\small =}&{\small -}\sum
\!\!\!\!\!\!\!\!\int^{\prime \prime }\sum \!\!\!\!\!\!\!\!\int^{\prime }%
\frac{(-)^{J-J^{\prime }}}{\widehat{J}}\left( \left. _{k}\left\langle
p^{\prime\prime \prime }q^{\prime\prime \prime }\alpha ^{\prime\prime \prime
}\right\Vert \overrightarrow{W}_{b}^{(k)}\left\Vert p^{\prime \prime
}q^{\prime \prime }\alpha ^{\prime \prime }\right\rangle _{k}\right. \right.
\label{new_C4} \\
&&\left. \times \left. _{k}\left\langle p^{\prime \prime }q^{\prime \prime
}\alpha ^{\prime \prime }\right\vert I_{0}\left\vert p^{\prime }q^{\prime
}\alpha ^{\prime }\right\rangle _{j}\right. \left. _{j}\left\langle
p^{\prime }q^{\prime }\alpha ^{\prime }\right\Vert \overrightarrow{W}%
_{e}^{(j)}\left\Vert pq\alpha \right\rangle _{j}\right. \left.
_{j}\left\langle pq\alpha \right. \left\vert \Psi \right\rangle \right.
\right)  \notag
\end{eqnarray}

This expression comprise only physical states and requires single basis
transformation operation. Evaluation of 3NF terms using old PWD of eq.(\ref%
{2pi_on_vec}) required two basis transformation operations: one $\left.
_{j}\left\langle p^{\prime \prime }q^{\prime \prime }\alpha ^{\prime \prime
}\right\vert I\left\vert p^{\prime }q^{\prime }\alpha ^{\prime
}\right\rangle _{i}\right. $ involving unphysical and physical states in both
ket and bra parts (this transformation is numerically at least four times heavier than
one coupling only physical states), whereas other $\left. _{k}\left\langle p^{\prime \prime
\prime \prime }q^{\prime \prime \prime \prime }\alpha ^{\prime \prime \prime
\prime }\right. \left\vert p^{\prime \prime \prime }q^{\prime \prime \prime
}\alpha ^{\prime \prime \prime }\right\rangle _{j}\right.$ transforming
combination of unphysical and physical ket states into the physical bra state
(this transformation is numerically at least two times heavier than
one coupling only physical states).
As mentioned in~\cite{Huber_3NF} the basis transformation operations consumes essentially
all the CPU time required to evaluate 3NF terms, while the time for
 the calculation of $_{i}\left\langle%
p^{\prime }q^{\prime }\alpha ^{\prime }\right\Vert W^{(i)}\left\Vert pq\alpha \right\rangle_{i}$ matrix elements is
negligible.
Therefore two-pion exchange 3NF terms can be evaluated by as much as six times
faster using eq.(\ref{Mat_form_me1}) \ and eq.(\ref{new_C4}) than using
old PWD.

In the end I should remark that presented method can be applied to evaluate
any momentum or coordinate space 3NF, which has a form: $V_{ijk}=\sum%
\limits_{cyc}V_{ij}V_{jk}$. Application of similar PWD technique to evaluate
UIX 3NF has been demonstrated in~\cite{Thesis}.

It is clear that described
method is not directly applicable for more complex 3NF, which involves more than two meson exchanges,
like Illinois force containing three-pion ring terms~\cite{Ill_3NF}. Nevertheless
variation of this technique is possible to evaluate also the terms of type $V_{ijk}=\sum%
\limits_{cyc}V_{ij}V_{jk}W_{ki}$ without introducing mixing with unphysical states.

\begin{appendix}
\section{\protect\bigskip}

Here we give explicit formulaes for the potential's terms $W$, required to
evaluate two-pion exchange 3NF. When working with Faddeev equations one
deals with binary particle interactions, which in proper Jacobi
coordinate set are expressed as the functions of momenta $p$ (and/or $p'$) and which are independent of momenta $q$ $(q')$.
  In contrary, our expressions for $W^{(j)}$%
-terms are the functions of $\overrightarrow{Q}_{j}=\overrightarrow{k}_{j}^{\prime }-%
\overrightarrow{k}_{j}\equiv\overrightarrow{q}_{j}^{\prime }-%
\overrightarrow{q}_{j}$ and they  are diagonal in Jacobi momentum $p_j$. In this sense
evaluation of $W$-terms are different from the evaluation of two-nucleon interaction terms, still
they are similar in numerical complexity.

The scalar potential term
$W_{a}^{(j)}=\widetilde{f}_{j}\left( \overrightarrow{\sigma }_{j}.%
\overrightarrow{Q}_{j}\right) $ is the simplest case to evaluate:
\begin{eqnarray}
\left. _{j}\left\langle p^{\prime }q^{\prime }\alpha ^{\prime }\right\vert
W_{a}^{(j)}\left\vert pq\alpha \right\rangle _{j}\right.  &=&\frac{\delta
(p^{\prime }-p)}{p^{\prime }p}\delta _{l_{p}l_{p}^{\prime }}\delta
_{j_{p}j_{p}^{\prime }}\delta _{s_{p}s_{p}^{\prime }}\delta
_{t_{p}t_{p}^{\prime }}\delta
_{j_{q}j_{q}^{\prime }}\delta _{JJ^{\prime }}\delta _{MM^{\prime }}\delta
_{TT^{\prime }}6\pi \sqrt{2\widehat{l}_{q}\widehat{l}_{q}^{\prime }}\left(
-\right) ^{j_{q}+l_{q}+\frac{3}{2}} \nonumber \\
&&\times \left\{
\begin{array}{ccc}
l_{q}^{\prime } & l_{q} & 1 \\
\frac{1}{2} & \frac{1}{2} & j_{q}%
\end{array}%
\right\} \sum_{\lambda }\sum_{k_{1}+k_{2}=1}\widehat{\lambda }\left(
\begin{array}{ccc}
k_{1} & \lambda  & l_{q}^{\prime } \\
0 & 0 & 0%
\end{array}%
\right) \left(
\begin{array}{ccc}
k_{2} & \lambda  & l_{q} \\
0 & 0 & 0%
\end{array}%
\right) \nonumber  \\
&&\times \left\{
\begin{array}{ccc}
k_{2} & l_{q} & \lambda  \\
l_{q}^{\prime } & k_{1} & 1%
\end{array}%
\right\} q^{\prime k_{1}}q^{k_{2}}g_{\lambda 2},
\end{eqnarray}%
with
\begin{equation}
g_{\lambda K}=\int_{-1}^{1}duP_{\lambda }(u)\frac{f(\widetilde{Q})}{%
Q^{2}+M_{\pi }^{2}}\frac{Q^{2}}{Q^{K}}.
\end{equation}

In the last expression $Q=\sqrt{q^{2}+q^{\prime 2}-2qq^{^{\prime }}u}$, $%
P_{\lambda }(u)$ is Legendre polynomial. The $f(\widetilde{Q})$ is model
defined nucleon high-momentum cut-off function as introduced in eq.(\ref%
{2pi_force}): in some models, as~\cite{TM_3NF,Navr_3NF}, cut-off is set on momentum transfer $\widetilde{Q}\equiv Q$;
in other models, as~\cite{Epel_3NF}, cut-off can be set on single particle momenta $\widetilde{Q}\equiv (q,q')$.
One should note that the operator $W_{a}^{(j)}$conserves the
three-nucleon angular momentum $J$, nevertheless it changes the parity of
the state to its opposite.

Next step is to evaluate reduced matrix elements of the vector potential $%
\overrightarrow{W}_{b}^{(j)}=\widetilde{f}_{j}\left( \overrightarrow{\sigma }%
_{j}.\overrightarrow{q}_{j}\right) \overrightarrow{q}_{j}$:
\begin{eqnarray}
\left. _{j}\left\langle p^{\prime }q^{\prime }\alpha ^{\prime }\right\Vert
\overrightarrow{W}_{b}^{(j)}\left\Vert pq\alpha \right\rangle _{j}\right.
&=&\frac{\delta (p^{\prime }-p)}{p^{\prime }p}\delta _{l_{p}l_{p}^{\prime
}}\delta _{j_{p}j_{p}^{\prime }}\delta _{s_{p}s_{p}^{\prime }}\delta
_{t_{p}t_{p}^{\prime }}\delta _{TT^{\prime }}2\pi \sum_{\lambda }\sum
_{\substack{ k_{1}+k_{2}=K \\ K=0,2}}\widehat{\lambda }\widehat{K}\nonumber  \\
&&\times (1+k_{1}k_{2})\sqrt{\frac{k_{1}!k_{2}!}{\widehat{k}_{1}!!\widehat{k}%
_{2}!!}}\sqrt{6\widehat{k}_{1}\widehat{k}_{2}\widehat{l}_{q}\widehat{l}%
_{q}^{\prime }\widehat{j}_{q}\widehat{j}_{q}^{\prime }\widehat{J}\widehat{J}%
^{\prime }}\left( -\right) ^{k_{2}+J^{\prime }+j_{q}+j_{p}} \nonumber \\
&&\times \left(
\begin{array}{ccc}
k_{1} & \lambda  & l_{q}^{\prime } \\
0 & 0 & 0%
\end{array}%
\right) \left(
\begin{array}{ccc}
k_{2} & \lambda  & l_{q} \\
0 & 0 & 0%
\end{array}%
\right) \left\{
\begin{array}{ccc}
j_{q} & j_{p} & J \\
J^{\prime } & 1 & j_{q}^{\prime }%
\end{array}%
\right\} \left\{
\begin{array}{ccc}
k_{2} & l_{q} & \lambda  \\
l_{q}^{\prime } & k_{1} & K%
\end{array}%
\right\}  \nonumber\\
&&\times \left\{
\begin{array}{ccc}
K & 1 & 1 \\
l_{q}^{\prime } & \frac{1}{2} & j_{q}^{\prime } \\
l_{q} & \frac{1}{2} & j_{q}%
\end{array}%
\right\} q^{\prime k_{1}}q^{k_{2}}g_{\lambda K},
\end{eqnarray}

 Finally, we present expressions for the reduced matrix elements of
the vector potentials $^{(k)}\overrightarrow{W}_{d}^{(j)} =%
\widetilde{f}_{i}\left( \overrightarrow{\sigma }_{j}\centerdot\overrightarrow{Q}%
_{j}\right) \left[ \overrightarrow{Q}_{j}\times \overrightarrow{\sigma }_{k}%
\right] $ and $\overrightarrow{W}_{e}^{(j)}=\widetilde{f}_{j}\left(
\overrightarrow{\sigma }_{j}\centerdot\overrightarrow{Q}_{j}\right) \left[
\overrightarrow{Q}_{j}\times (2\overrightarrow{S}-\overrightarrow{\sigma }_{j})\right] $:
\begin{eqnarray}
\left. _{j}\left\langle p^{\prime }q^{\prime }\alpha ^{\prime }\right\Vert
^{(k)}\overrightarrow{W}_{d}^{(j)}\left\Vert pq\alpha \right\rangle
_{j}\right.  &=&\frac{\delta (p^{\prime }-p)}{p^{\prime }p}\delta
_{l_{p}l_{p}^{\prime }}\delta _{t_{p}t_{p}^{\prime }}\delta _{TT^{\prime
}}12\pi i\sum_{\lambda LL^{\prime }SS^{\prime }h}\sum_{\substack{ %
k_{1}+k_{2}=K \\ K=0,2}}\widehat{\lambda }\widehat{h}\widehat{K}\widehat{L}\widehat{L}%
^{\prime }\widehat{S}\widehat{S}^{\prime }  \notag \\
&&\times (1+k_{1}k_{2})\sqrt{\frac{k_{1}!k_{2}!}{\widehat{k}_{1}!!\widehat{k}%
_{2}!!}}\sqrt{6\widehat{k}_{1}\widehat{k}_{2}\widehat{j}_{p}\widehat{j}%
_{p}^{\prime }\widehat{j}_{q}\widehat{j}_{q}^{\prime }\widehat{l}_{q}%
\widehat{l}_{q}^{\prime }\widehat{J}\widehat{J}^{\prime }\widehat{s}_{p}%
\widehat{s}_{p}^{\prime }}\left( -\right) ^{L^{\prime
}+k_{2}+l_{q}+l_{p}+h+1+s_{p}^{\prime }}  \notag \\
&&\times \left(
\begin{array}{ccc}
k_{1} & \lambda  & l_{q}^{\prime } \\
0 & 0 & 0%
\end{array}%
\right) \left(
\begin{array}{ccc}
k_{2} & \lambda  & l_{q} \\
0 & 0 & 0%
\end{array}%
\right) \left\{
\begin{array}{ccc}
l_{q} & l_{p} & L \\
L^{\prime } & K & l_{q}^{\prime }%
\end{array}%
\right\} \left\{
\begin{array}{ccc}
h & K & 1 \\
1 & 1 & 1%
\end{array}%
\right\}   \notag \\
&&\times \left\{
\begin{array}{ccc}
k_{2} & l_{q} & \lambda  \\
l_{q}^{\prime } & k_{1} & K%
\end{array}%
\right\} \left\{
\begin{array}{ccc}
\frac{1}{2} & \frac{1}{2} & s_{p}^{\prime } \\
s_{p} & 1 & \frac{1}{2}%
\end{array}%
\right\} \left\{
\begin{array}{ccc}
l_{p} & s_{p} & j_{p} \\
l_{q} & \frac{1}{2} & j_{q} \\
L & S & J%
\end{array}%
\right\} \left\{
\begin{array}{ccc}
l_{p}^{\prime } & s_{p}^{\prime } & j_{p}^{\prime } \\
l_{q}^{\prime } & \frac{1}{2} & j_{q}^{\prime } \\
L^{\prime } & S^{\prime } & J^{\prime }%
\end{array}%
\right\}   \notag \\
&&\times \left\{
\begin{array}{ccc}
K & h & 1 \\
L^{\prime } & S^{\prime } & J^{\prime } \\
L & S & J%
\end{array}%
\right\} \left\{
\begin{array}{ccc}
1 & 1 & h \\
s_{p}^{\prime } & \frac{1}{2} & S^{\prime } \\
s_{p} & \frac{1}{2} & S%
\end{array}%
\right\} q^{\prime k_{1}}q^{k_{2}}g_{\lambda K}
\end{eqnarray}%
and
\begin{eqnarray}
\left. _{j}\left\langle p^{\prime }q^{\prime }\alpha ^{\prime }\right\Vert
\overrightarrow{W}_{e}^{(j)}\left\Vert pq\alpha \right\rangle _{j}\right. &=&%
\frac{\delta (p^{\prime }-p)}{p^{\prime }p}\delta _{s_{p}1}\delta
_{s_{p}s_{p}^{\prime }}\delta _{l_{p}l_{p}^{\prime }}\delta
_{t_{p}t_{p}^{\prime }}\delta _{TT^{\prime }}24\pi i\sum_{\lambda LL^{\prime
}SS^{\prime }h}\sum_{\substack{ k_{1}+k_{2}=K  \\ K=0,2}}\widehat{\lambda }\widehat{h}%
\widehat{K}\widehat{L}\widehat{L}^{\prime }\widehat{S}\widehat{S}%
^{\prime }  \notag \\
&&\times (1+k_{1}k_{2})\sqrt{\frac{k_{1}!k_{2}!}{\widehat{k}_{1}!!\widehat{k}%
_{2}!!}}\sqrt{6\widehat{k}_{1}\widehat{k}_{2}\widehat{j}_{p}\widehat{j}%
_{p}^{\prime }\widehat{j}_{q}\widehat{j}_{q}^{\prime }\widehat{l}_{q}%
\widehat{l}_{q}^{\prime }\widehat{J}\widehat{J}^{\prime }}\left( -\right)
^{L^{\prime }+k_{2}+l_{q}+l_{p}+h+1}  \nonumber \\
&&\times \left(
\begin{array}{ccc}
k_{1} & \lambda & l_{q}^{\prime } \\
0 & 0 & 0%
\end{array}%
\right) \left(
\begin{array}{ccc}
k_{2} & \lambda & l_{q} \\
0 & 0 & 0%
\end{array}%
\right) \left\{
\begin{array}{ccc}
l_{q} & l_{p} & L \\
L^{\prime } & K & l_{q}^{\prime }%
\end{array}%
\right\} \left\{
\begin{array}{ccc}
h & K & 1 \\
1 & 1 & 1%
\end{array}%
\right\} \nonumber\\
&&\times \left\{
\begin{array}{ccc}
k_{2} & l_{q} & \lambda \\
l_{q}^{\prime } & k_{1} & K%
\end{array}%
\right\} \left\{
\begin{array}{ccc}
l_{p} & s_{p} & j_{p} \\
l_{q} & \frac{1}{2} & j_{q} \\
L & S & J%
\end{array}%
\right\} \left\{
\begin{array}{ccc}
l_{p}^{\prime } & s_{p}^{\prime } & j_{p}^{\prime } \\
l_{q}^{\prime } & \frac{1}{2} & j_{q}^{\prime } \\
L^{\prime } & S^{\prime } & J^{\prime }%
\end{array}%
\right\}\nonumber \\
&&\times \left\{
\begin{array}{ccc}
K & h & 1 \\
L^{\prime } & S^{\prime } & J^{\prime } \\
L & S & J%
\end{array}%
\right\} \left\{
\begin{array}{ccc}
1 & 1 & h \\
s_{p}^{\prime } & \frac{1}{2} & S^{\prime } \\
s_{p} & \frac{1}{2} & S%
\end{array}%
\right\} q^{\prime k_{1}}q^{k_{2}}g_{\lambda K}.
\end{eqnarray}

The $\overrightarrow{W}_{b}^{(j)}$, $^{(k)}\overrightarrow{W}_{d}^{(j)}$ and
$\overrightarrow{W}_{e}^{(j)}$potential terms conserve the parity of
three-nucleon state, however it may change its total angular momentum.

Using simple tensor algebra relations, one has:
\begin{equation*}
\left( \overrightarrow{\sigma }_{i}\centerdot \overrightarrow{Q}_{i}\right) %
\left[ \overrightarrow{Q}_{i}\times \overrightarrow{\sigma }_{i}\right]
=3i\sum_{gh}\sqrt{2\widehat{g}\widehat{h}}\left\{
\begin{array}{ccc}
1 & 1 & 0 \\
1 & 1 & 1 \\
g & h & 1%
\end{array}%
\right\} \left\{ \left\{ \overrightarrow{Q}_{i}\otimes \overrightarrow{Q}%
_{i}\right\} _{g}\otimes \left\{ \overrightarrow{\sigma }_{i}\otimes
\overrightarrow{\sigma }_{i}\right\} _{h}\right\} _{1}.
\end{equation*}
In this equation the product $\left\{ \overrightarrow{\sigma }_{i}\otimes
\overrightarrow{\sigma }_{i}\right\} _{h}$ acts on spin one-half states and
it is equal zero unless $h=0$. However $g=1$, if $h=0$ and $\left\{
\overrightarrow{Q}_{i}\otimes \overrightarrow{Q}_{i}\right\} _{g=1}\equiv 0$
as a vector product of two parallel vectors vanishes. Therefore the last
equation is identically zero and  $\overrightarrow{W}_{f}^{(i)}\equiv 0.$

PWD basis transformation operations $\left.
_{j}\left\langle p^{\prime \prime }q^{\prime \prime }\alpha ^{\prime \prime
}\right\vert I\left\vert p^{\prime }q^{\prime }\alpha ^{\prime
}\right\rangle _{i}\right. $ have been discussed and necessary expressions
have been given in ref.~\cite{Huber_3NF,Huber_3NF_2}.
\end{appendix}

\begin{acknowledgments}
I would like to thank my former Ph.D. supervisor Claude Gignoux. In fact, idea
of the PWD presented in this work have born already long time ago during discussion
with Claude.
\end{acknowledgments}


\begin{thebibliography}{9}
\bibitem{Huber_3NF} D. H\"{u}ber, H. Wita\l a, A. Nogga et al., Few-Body\
Syst. \textbf{22} (1997) 107.

\bibitem{3NF_gen} AIP Conference Proceedings \textbf{1011}, edited by H.
Sakai, K. Sekiguchi, B.F. Gibson (AIP, 2008);
http://proceedings.aip.org/dbt/dbt.jsp?KEY=APCPCS.

\bibitem{Epel_3NF} E. Epelbaum, A. Nogga, W. Gl\"ockle et al., Phys. Rev.
\textbf{C66} (2002) 064001.

\bibitem{TM_3NF} J. L. Friar, D. H\"{u}ber, and U. van Kolck, Phys. Rev.
\textbf{C 59} (1999) 53.

\bibitem{Navr_3NF} P. Navratil, Few Body Syst. \textbf{41} (2007) 117.

\bibitem{Thesis} R. Lazauskas, PhD Thesis, Universit\'e Joseph Fourier,
Grenoble (2003).

\bibitem{Huber_3NF_2} D. H\"{u}ber, J.L. Friar, A. Nogga et al., Few-Body\
Syst. \textbf{30} (2001) 95.

\bibitem{Ill_3NF} S.C. Pieper, V.R. Pandharipande et al., Phys. Rev.
\textbf{C64} (2001) 014001.
\end{thebibliography}
\end{document}